\documentclass[12pt]{amsart}

\theoremstyle{plain}

\newtheorem{theorem}{Theorem}[section]

\newtheorem{lemma}[theorem]{Lemma}

\theoremstyle{definition}
\theoremstyle{definition}

\newcommand{\be}{\begin{equation}}
\newcommand{\ee}{\end{equation}}
\newcommand{\bea}{\begin{eqnarray}}
\newcommand{\eea}{\end{eqnarray}}
\newcommand{\bean}{\begin{eqnarray}}
\newcommand{\eean}{\end{eqnarray}}
\newcommand{\ba}{\begin{array}}
\newcommand{\ea}{\end{array}}
\newcommand{\ep}{\epsilon}

\newcommand{\ga}{\gamma}

\newcommand{\la}{\lambda}

\newcommand{\pa}{\partial}

\newcommand{\no}{\nonumber}
\newcommand{\eq}{\eqref}
\begin{document}

 \title[waterbag model(II)]
{On the water-bag model of dispersionless KP hierarchy  (II)}

\author[Jen-Hsu Chang]
{Jen-Hsu Chang \\
Department of Computer Science\\
 National Defense University\\
 Tauyuan, Taiwan\\
 E-mail: jhchang@ccit.edu.tw}
\maketitle

\begin{abstract}
We construct the bi-Hamiltonian structure of the waterbag model of dKP and
establish the third-order  Hamiltonian
operator associated with the waterbag model. Also, the  symmetries and conserved
densities of rational type  are discussed.
\end{abstract}
\section{Introduction}
The dispersionless KP hierarchy(dKP or Benney moment chain) is  defined by
\begin{equation}
\pa_{t_n}\la(z)=\{\la(z), B_n(z)\},\quad n=1,2, \cdots., \label{kpp}
\end{equation}
where the Lax operator $\la(z)$ is
\begin{equation}
\la(z)=z+\sum_1^{\infty} v_{n+1} z^{-n}  \label{pl}
\end{equation}
and
\[B_n(z)=\frac{[\la^n(z)]_{+}}{n},\quad i=1,2,3, \cdots  \quad t_1=x. \]
Here $[\cdots]_{+}$  denotes the non-negative part of the Laurent series $\la^n(z)$.
For example,
\[B_2=\frac{z^2}{2}+v_2,\qquad B_3=\frac{z^3}{3}+v_2 z+v_3.\]
Finally,
the bracket  in  \eqref{kpp} denotes the natural Poisson bracket on the space of functions of
the two variables $(x,z)$:
\begin{equation}
\{f(x,z), g(x,z)\}=\pa_x f \pa_z g-\pa_x g \pa_z f. \label{po}
\end{equation}
The compatibility of  \eqref{kpp}  coincides with the zero-curvature equation
\begin{equation}
\pa_m B_n (z)-\pa_n B_m (z)=\{B_n (z),B_m(z)\}. \label{ze}
\end{equation}
If we denote $t_2=y$ and $t_3=t$, then the equation \eqref{ze} for $m=2, n=3$ gives
\begin{eqnarray*}
v_{3x}&=&v_{2y} \\
v_{3y}&=& v_{2t}-v_2v_{2x},
\end{eqnarray*}
from which the dKP equation is derived($v_2=v$)
\begin{equation}
v_{yy}=(v_t-vv_x)_x.\label{dkp}
\end{equation}
According to the  dKP theory \cite{Ak,KG, kr2, Ta1}, there exists a wave function $S(\la,x,t_2,t_3, \cdots)$ such that $z=S_x$ and satisfies
the Hamiltonian-Jacobian equation
\begin{equation}
\frac{\pa S}{\pa t_n}=B_n(z)\vert_{z=S_x}. \label{hj}
\end{equation}
It can be seen that the compatibility of \eq{hj} also  implies the zero-curvature equation \eqref{ze}.
 Now, we expand $B_n(z)$ as
\[B_n(z(\lambda))=\frac{[\la^n(z)]_+}{n}=\frac{\la^n}{n}-\sum_{i=0}^{\infty}G_{in}\la^{-i-1},\]
where the coefficients can be calculated by the residue form:
\[G_{in}=-res_{\la=\infty}(\la^i B_{n}(z)d\la)=\frac{1}{i+1} res_{z=\infty}(\la^{i+1} \frac{\pa B_n(z)}{\pa z}dz),\]
from which the symmetry property
\[G_{in}=G_{ni}\]
can be easily deduced.
Moreover \cite{Ta1}, it can be shown that the polynomials $B_n$ must
 satisfy the integrability condition
\[\frac{\pa B_m(\la)}{\pa t_n}=\frac{\pa B_n(\la)}{\pa t_m},\]
from which in turns follows the integrability of the coefficients $G_{in}$, i.e., there exists
 the free energy $\mathcal{F}$ (dispersionless $\tau$ function) such that
\[G_{in}=\frac{\pa^2 \mathcal{F}}{\pa t_i \pa t_n}.\]
This latter function may be for example used to  invert formula  \eq{pl} :
\begin{equation}
z=\la-\frac{\mathcal{F}_{11}}{\la}-\frac{\mathcal{F}_{12}}{2\la^2}-\frac{\mathcal{F}_{13}}{3\la^3}-
\frac{\mathcal{F}_{14}}{4\la^4}-\cdots, \label{in}
\end{equation}
where $\mathcal{F}_{1n}$ are polynomials of $v_2,v_3,\cdots,v_{n+1}$ and in fact
\begin{equation}
h_n \equiv \frac{\mathcal{F}_{1n}}{n}=res_{z=\infty} \frac{\lambda^n}{n}dz \label{co}
\end{equation}
are  the conserved densities for the dKP hierarchy \eqref{kpp}.
In \cite{bm,Car2}, it's proved that dKP hierarchy \eqref{kpp} is equivalent
to the dispersionless Hirota equation
\begin{equation}
D(\la)S(\la^{'}) =-log  \frac{z(\la)-z(\la^{'})}{\la}, \label{hr}
\end{equation}
where $D(\la)$ is the operator
$\sum_{n=1}^{\infty} \frac{1}{n\la^n}\frac{\pa}{\pa t_n}$. \\
\indent Next, we consider the symmetry constraint \cite{bk}
\begin{equation}
\mathcal{F}_x=\sum_{i=1}^M \ep_i S_i,\label{sy}
\end{equation}
where $S_i=S(\la_i)$, $\la_i$ are points in the complex plane and $\ep_i$ are constants.
Notice that from \eqref{in} we know
\[ D(\la)\mathcal{F}_x=\la -z. \]
On the other hand, by \eqref{hr} and \eqref{sy}, we also have
\begin{eqnarray*}
 D(\la)\mathcal{F}_x &=& \sum_{i=1}^M \ep_i D(\la) S_i=-\sum_{i=1}^M \ep_i
\ln \frac{z (\la)-z(\la_i)}{\la} \\
= &-& \sum_{i=1}^M \ep_i \ln (z-p^i)+ (\sum_{i=1}^M \ep_i) \ln \la,
\end{eqnarray*}
where $z=z(\la)$ and $p^i=z(\la_i)$. We assume that
\[\sum_{i=1}^M \ep_i=0 \]
and then we get
the waterbag reduction \cite{bk,pa1}
\begin{eqnarray}
\la &=& z-\sum_{i=1}^M \ep_i \ln (z-p^i)  \label{wat} \\
&=& z + \sum_{n=1}^{\infty} \frac{ v_{n+1}}{z^n},
\end{eqnarray}
where $v_{n+1}=\frac{1}{n} \sum_{i=1}^M \ep_i( p^i)^n $.
From this  we obtain
\[B_2(z)=\frac{1}{2}z^2+ \sum_{i=1}^M \ep_i p^i. \]
So ($t_2=y$)
\begin{equation}
\pa_y p^i= \pa_x[ \frac{1}{2}(p^i)^2 +\sum_{i=1}^M \ep_i p^i] \label{pr} \\
\end{equation}
In \cite{ch}, the two-component case of \eqref{pr} is investgated. We see that the
equation \eqref{pr} can be written  as  the Hamiltonian system
\begin{equation}
 \left[\begin{array}{c}
p^1 \\
p^2 \\
\vdots \\
 p^M
\end{array} \right]_y
=\frac{1}{2}\left[\begin{array}{ccccc}
\frac{1}{\ep_1} & 0 & \ldots & \ldots  & 0 \\
0 & \frac{1}{\ep_2} & 0 & \ldots & 0 \\
\vdots & \vdots & \ddots & \vdots & 0  \\
0 & 0 & \ldots & 0 & \frac{1}{\ep_M}
\end{array} \right] \pa_{x}
\left[\begin{array}{c}
\frac{\delta H_3}{\delta  p^1} \\
\frac{\delta H_3}{\delta  p^2} \\
\vdots \\
\frac{\delta H_3}{\delta p^M}
\end{array} \right],  \no
\end{equation}
where $\delta$ is the variation derivative and
\begin{eqnarray*}
H_3 &=& \frac{1}{3} \int dx Res(\la^3 dz) =\int dx (v_2^2+v_4) \\
&=& \int dx [ (\sum_{i=1}^M \ep_i p^i)^2+ \frac{1}{3}(\sum_{i=1}^M \ep_i (p^i)^3)]
\end{eqnarray*}
A bi-Hamiltonian structure is defined as (in  the case of dKP)
 \begin{equation}
 \left[\begin{array}{c}
p^1 \\
p^2 \\
\vdots \\
 p^M
\end{array} \right]_y
= -\frac{1}{2} J_1
\left[\begin{array}{c}
\frac{\delta H_3}{\delta  p^1} \\
\frac{\delta H_3}{\delta  p^2}\\
\vdots \\
\frac{\delta H_3}{\delta  p^M}
\end{array} \right]  =J_2
\left[\begin{array}{c}
\frac{\delta H}{\delta  p^1} \\
\frac{\delta H}{\delta  p^2}\\
\vdots \\
\frac{\delta H}{\delta  p^M}
\end{array} \right], \no
\end{equation}
where
\[ J_1= -\left[\begin{array}{ccccc}
\frac{1}{\ep_1 } & 0 & \ldots & \ldots  & 0 \\
0 & \frac{1}{\ep_2} & 0 & \ldots & 0 \\
\vdots & \vdots & \ddots & \vdots & 0  \\
0 & 0 & \ldots & 0 & \frac{1}{\ep_M}
\end{array} \right] \pa_{x}     \]
and $J_2$ is also a Hamiltonian operator, which is
compatible with $J_1$, i.e., $J_1+c J_2$
is also a Hamiltonian one for any complex number $c$ \cite{du1,Ma,Li}. We hope to find
 $J_2$ and
the related Hamiltonian $H$.\\
\indent Furthermore, from the bi-Hamiltonian structure  \eqref{bi} or \eqref{bi2}(see below) of the  waterbag
model \eqref{wat}, we also find the recursion operator $\hat R$ in \eqref{st}(see below) is local.
 Then, according to the bi-Hamiltonian theory \cite{Ma,Pe}, one can construct rational
symmetries using the local recursion operator $\hat R$(equation \eqref{st}).  Hence the higher-order rational conserved
densities (quasi-rational functions) are investigated. \\
\indent  This paper is organized as follows. In the next section, we construct the
the  bi-Hamiltonian structure of the waterbag model from the Landau-Ginsburg formulation
in topological field theory. Section 3 is devoted to investigating  the quasi-rational
symmetries and the corresponding conserved densities of the waterbag model.
 In the final section, one discusses some problems to be investigated

\section{Free energy and bi-Hamiltonian structure}
In this section, we investigate the relations between bi-Hamiltonian structure
and free energy of waterbag model.\\
\indent The free energy is a function $\mathbb{F}(t^1, t^2, \cdots, t^n)$ such that
 the associated functions,
\[c_{ijk}= \frac{\pa^3 \mathbb{F}}{ \pa t^i \pa t^j \pa t^k},\]
satisfy the following conditions.
\begin{itemize}
\item The matrix $\eta_{ij}=c_{1ij}$ is constant and non-degenerate.
 This together with the inverse matrix $\eta^{ij}$ are used to raise
 and lower indices.
\item The functions $c_{jk}^{i}=\eta^{ir}c_{rjk}$ define an associative
 commutative algebra with a unity element(Frobenius algebra).
\end{itemize}
\indent Equations of associativity give  a system of non-linear PDE
for $\mathbb{F}(t)$
\[\frac{\pa^3 \mathbb{F}(t)}{\pa t^{\alpha} \pa t^{\beta}
 \pa t^{\lambda}} \eta^{\lambda \mu} \frac{\pa^3 \mathbb{F}(t)}
 {\pa t^{\mu} \pa t^{\ga} \pa t^{\sigma}}  =
\frac{\pa^3 \mathbb{F}(t)}{\pa t^{\alpha} \pa t^{\ga}
 \pa t^{\lambda}} \eta^{\lambda \mu} \frac{\pa^3 \mathbb{F}(t)}
 {\pa t^{\mu} \pa t^{\beta} \pa t^{\sigma}}. \]
These equations constitute the Witten-Dijkgraaf-Verlinde-Verlinde (or WDVV)
equations. The geometrical setting in which to understand the free energy $\mathbb{F}(t)$
is the Frobenius manifold \cite{du1,Du4}. One way to construct such manifold is derived via
Landau-Ginzburg formalism as the structure on the parameter space $M$ of the appropriate
form
\[ \la=\la(z;t^1,t^2, \cdots, t^n).\]
The Frobenius structure is given by the flat metric
\[\eta( \pa, \pa')=-\sum res_{d \la =0} \{\frac{\pa(\la dz)\pa'(\la dz)}{ d \la (z)} \} \]
and the tensor
\[c( \pa, \pa', \pa^{''})=-\sum res_{d \la =0} \{\frac{\pa(\la dz)\pa'(\la dz)\pa^{''}(\la dz)}
{ d \la (z) dz}  \} \]
defines a totally symmetric $(3,0)$-tensor $c_{ijk}$. \\
\indent Geometrically, a solution of WDVV equation defines a multiplication
\[ \circ: TM \times TM \longrightarrow TM \]
of vector fields on the parameter space $M$, i.e,
\[ \pa_{t^{\alpha}} \circ \pa_{t^{\beta}}= c_{\alpha \beta}^{\gamma}(t) \pa_{t^{\gamma}}.\]
From $c_{\alpha \beta}^{\gamma}(t)$, one can construct intrgable hierarchies whose corresponding
Hamiltonian densities are defined recursively by the formula
\bean
\frac{\pa^2 \psi_{\alpha}^{(l)}}{\pa t^i \pa t^j}=c_{ij}^k \frac{\pa \psi_{\alpha}^{(l-1)}}
{\pa t^k}, \label{rec}
\eean
where $l \geq 1, \alpha =1,2, \cdots, n,$ and $ \psi_{\alpha}^{0}=\eta_{\alpha
\epsilon}t^{\ep}$. The integrability conditions for this systems are automatically
satisfied when the $c_{ij}^k$ are defined as above.\\
\indent For the waterbag model \eqref{wat}, we have the following
\begin{theorem}\cite{fs}:
\begin{eqnarray*}
\eta(\frac{\pa}{\pa p^i}, \frac{\pa}{\pa p^j})&=&\eta_{ij}=-\ep_i \delta_{i,j} \quad i,j=1,2, \cdots M;\\
c(\frac{\pa}{\pa p^{\alpha}}, \frac{\pa}{\pa p^{ \beta}},\frac{\pa}{\pa p^{ \gamma}} ) &=&
c_{\alpha\beta\gamma}=0, \quad \alpha, \quad \beta, \quad \gamma \quad distinct ;\\
c(\frac{\pa}{\pa p^{\alpha}}, \frac{\pa}{\pa p^{ \alpha}},\frac{\pa}{\pa p^{ \beta}} ) &=&
c_{\alpha\alpha\beta}=\frac{\ep_{\alpha} \ep_{\alpha}}{p^{\beta}-p^{\alpha}},
\quad \alpha  \neq \beta  ; \\
c(\frac{\pa}{\pa p^{\alpha}}, \frac{\pa}{\pa p^{ \alpha}},\frac{\pa}{\pa p^{ \alpha}} ) &=&
c_{\alpha\alpha\alpha}=-\ep_{\alpha}+\sum_{r\neq \alpha}\frac{\ep_{\alpha} \ep_{r}}
{p^{\alpha}-p^r},
\quad \alpha  \neq \beta  . \\
\end{eqnarray*}
\end{theorem}
\noindent Let's define
\begin{equation}
\Omega=\sum_{i=1}^M \frac{\pa}{\pa p^i} \label{id}
\end{equation}
Then we can see that
\begin{equation}
\eta(\frac{\pa}{\pa p^i}, \frac{\pa}{\pa p^j})=
c(\frac{\pa}{\pa p^i}, \frac{\pa}{\pa p^j},\Omega). \label{fla}
\end{equation}
\indent From the theorem, one can get the free energy associated with the waterbag model \eqref{wat},
noting that $p^i$ are flat coordinates,
\begin{equation}
\mathbb{F}(\vec{p})=-\frac{1}{6} \sum_{k=1}^M \ep_k (p^k)^3+\frac{1}{8} \sum_{i \neq j}
\ep_i \ep_j (p^i-p^j)^2 \ln (p^i-p^j)^2, \label{fre}
\end{equation}
where $\vec{p}=(p^1, p^2, \cdots,p^M)$ and from \eqref{fla} one has
\[t^1=\sum_{i=1}^M p^i. \]
Also, we have
\begin{eqnarray*}
c_{\beta\gamma}^{\alpha} &=& 0,\quad \alpha, \beta, \gamma \quad distinct ;\\
c_{\alpha\alpha}^{\beta} &=& \frac{\ep_{\alpha}}{p^{\alpha}-p^{\beta}}, \quad \alpha  \neq \beta \\
c_{\alpha\beta}^{\alpha} &=& \frac{\ep_{\beta}}{p^{\alpha}-p^{\beta}}, \quad \alpha  \neq \beta \\
c_{\alpha\alpha}^{\alpha} &=&1-\sum_{r \neq \alpha} \frac{\ep_{r}}{p^{\alpha}-p^r}.
\end{eqnarray*}
If we define $e_i=\frac{\pa \la}{\pa p^i}=
\frac{\ep_i}{z-p^i}$, then it's not difficult to see that
\begin{equation}
e_ie_j=c_{ij}^ke_k+Q_{ij}(\frac{d \la}{dz}), \label{as}
\end{equation}
where
\[Q_{ij}=\left\{
\begin{array}{ll}-\frac{\ep_i}{z-p^i}, & i=j,\\ 0, & i \neq j.\end{array} \right. \]
Hence we have the recursion relation
\begin{eqnarray}
\frac{\pa^2 h_n}{\pa p^i \pa p^j} &=& \frac{\pa}{\pa p^i}
\oint_{\infty} \la^{n-1} \frac{\pa \la}{\pa p^j}dz= \frac{\pa}{\pa p^i}
\oint_{\infty} \la^{n-1} \frac{\ep_j}{z- p^j}dz  \no \\
&=& (n-1) \oint_{\infty} \la^{n-2} \frac{\ep_i}{z- p^i} \frac{\ep_j}{z- p^j} dz
+\oint_{\infty} \la^{n-1} \frac{\pa}{\pa p^i}( \frac{\ep_j}{z- p^j}) dz \no \\
&=& (n-1) \oint_{\infty} \la^{n-2}( c^k_{ij} e_k +Q_{ij}) \frac{d \la}{dz}-\oint_{\infty}
\la^{n-1}\frac{\pa}{\pa z}( \frac{\ep_j}{z- p^j}) dz  \no \\
&=& c^k_{ij} (n-1)\oint_{\infty} \la^{n-2} \frac{\ep_k}{z- p^k}  dz-\oint_{\infty}
\frac{\pa}{\pa z}( \la^{n-1}\frac{\ep_j}{z- p^j}) dz \no \\
&=& c^k_{ij} \frac{\pa}{\pa p^k} \oint_{\infty} \la^{n-1} dz \no \\
&=& (n-1) c^k_{ij} \frac{\pa h_{n-1}}{\pa p^k} , \quad  n \geq  1. \label{rec}
\end{eqnarray}
Morever, we also have
\begin{eqnarray}
(\sum_i \pa_{p^i})h_n &=& \oint_{\infty} \la^{n-1} (\sum_{i=1}^M  \frac{\ep_i}{z-p^i}) dz \no \\
&=& \oint_{\infty} \la^{n-1} (1-\frac{d \la}{dz})=(n-1) h_{n-1}. \label{rec2}
\end{eqnarray}
We can express \eqref{rec} as
\[\frac{\pa h_n}{\pa p^i}= (n-1) \eta^{kl} \pa_x^{-1}\left[ (\frac{\pa^2 \mathbb{F}}
{\pa p^i \pa p^l})_x \frac{\pa h_{n-1}}{\pa p^k}\right], \]
where
\[ \eta^{kl}= -\left[\begin{array}{ccccc}
\frac{1}{\ep_1} & 0 & \ldots & \ldots  & 0 \\
0 & \frac{1}{\ep_2} & 0 & \ldots & 0 \\
\vdots & \vdots & \ddots & \vdots & 0  \\
0 & 0 & \ldots & 0 & \frac{1}{\ep_M}
\end{array} \right]      \]

Therefore we obtain the recursion operator \cite{pa2}
\begin{eqnarray}
R_i^k &=& \eta^{kl}\pa_x^{-1} \left[(\frac{\pa^2 \mathbb{F}}{\pa p^i \pa p^l})_x \right]
=\pa_x^{-1}\left[ \eta^{kl}(\frac{\pa^2 \mathbb{F}}{\pa p^i \pa p^l})_x \right] \no \\
&=& \pa_x^{-1} W_{ix}^k= \pa_x^{-1} \eta^{kl}\frac{\pa^3 \mathbb{F}}{\pa p^i \pa p^l
 \pa p^{\varepsilon}} p^{\varepsilon}_x ,   \label{op}
\end{eqnarray}
where
\begin{equation}
\label{wate} \\
\end{equation}
\begin{eqnarray}
&& W_i^k = \eta^{kl}\frac{\pa^2 \mathbb{F}}{\pa p^i \pa p^l} = \no  \\
&& \left[\begin{array}{cccc}
p^1- \sum_{l \neq 1}\ep_l \ln (p^1-p^l)  & \ep_2 \ln(p^1-p^2) & \ldots  & \ep_M \ln(p^1-p^M) \\
 \ep_1 \ln(p^2-p^1)& p^2- \sum_{l \neq 2}\ep_l \ln (p^2-p^l) &  \ldots & \ep_M \ln(p^2-p^M) \\
\vdots & \vdots  & \vdots & \vdots \\
 \ep_1 \ln(p^M-p^1)& \ep_2 \ln(p^M-p^2) & \ldots & p^M- \sum_{l \neq M}\ep_l \ln (p^M-p^l)
 \end{array} \right]  \no
\end{eqnarray}
Then
\begin{equation}
\left[\begin{array}{c}
\frac{\pa h_{n} }{\pa  p^1} \\
\frac{\pa h_{n} }{\pa  p^2}\\
\vdots \\
\frac{\pa h_{n} }{\pa  p^M}
\end{array} \right]=(n-1)(n-2) R^2
\left[\begin{array}{c}
\frac{\pa h_{n-2} }{\pa  p^1} \\
\frac{\pa h_{n-2} }{\pa  p^2}\\
\vdots \\
\frac{\pa h_{n-2} }{\pa  p^M}
\end{array} \right]. \label{rec3}
\end{equation}
Also, it's known that the Hamiltonian operators \cite{Mo2,Mo3}(see also \cite{pa2})
\begin{eqnarray*}
 J_1 &=& \eta^{ij} \pa_x \\
 J_2 &=& \sum_{m=1}^M \sum_{\alpha =1}^M \eta^{m \alpha}\eta^{i \ep}\eta^{jr}
\frac{\pa^3 \mathbb{F}}{\pa p^{\ep} \pa p^{m} \pa p^{k}} p^k_x \pa^{-1}_x
\frac{\pa^3 \mathbb{F}}{\pa p^{\alpha} \pa p^r \pa p^{s}} p^s_x \\
     &=& \sum_{m=1}^M \sum_{\alpha =1}^M (W_m^i)_x \pa^{-1}_x (W_{\alpha}^j)_x
\end{eqnarray*}
 are compatible. Consequently, using \eqref{rec3} or $R^2=J_1^{-1} J_2$
 , we obtain the bi-Hamiltonian structure for the waterbag model \eqref{wat}, $n \geq 2$,
 \begin{eqnarray}
\pa_{t_n} p^l &=& \frac{-1}{n} \eta^{li} \pa_x \frac{\pa h_{n+1}}{\pa p^i} \no \\
&=& -(n-1) \eta^{m \alpha}(W_m^l)_x \pa_x^{-1} (W_{\alpha}^i)_x \frac{\pa h_{n-1}}{\pa p^i},
\label{bi}
\end{eqnarray}
where $W$ is defined by \eqref{wate}. \\
\indent For $n=2$, i.e.\eqref{pr}, one can directly verify the
 bi-Hamiltonian structure \eqref{bi}.\\
$\mathbf{Remark:}$ Using the Legendre-type transformation for WDVV equation in \cite{du1},
 we can introduce
new flat coordinates from the first row vector of \eqref{wate}
\begin{eqnarray*}
a_1 &=& \frac{1}{\ep_1} [ p^1- \sum_{l \neq 1}\ep_l \ln (p^1-p^l)]  \\
a_k &=&  \ln (p^1-p^k), \quad k=2, 3,  4, \cdots,  M.
\end{eqnarray*}
The inverse transformation of above equation is
\begin{eqnarray*}
p^1 &=& \sum_{i=1}^M \ep_i a_i   \\
p^k &=& \sum_{i=1}^M \ep_i a_i-e^{a_k},  \quad k=2, 3, 4, \cdots,  M.
\end{eqnarray*}
Then the new free energy satisfying the WDVV equation is \cite{pa1}
\begin{eqnarray*}
 F &=& \frac{ \ep_1^2 (a_1)^3}{6}+ \frac{\ep_1 a_1}{2} \sum_{m \neq 1} \ep_m (a_m)^2
 +P_3 (\vec{ a}) - \sum_{m \neq 1} \ep_m e^{a_m}  \\
 &+& \frac{1}{2}\sum_{ 1 <m < k} \ep_m \ep_k
 [ Li_3(e^{ a_k-a_m} )+Li_3(e^{ a_m-a_k} )],
 \end{eqnarray*}
where
\[P_3 (\vec{ a})= \sum_{m \neq 1} \frac{\ep_m (\ep_m-\ep_1)( a_m^3) }{6} +
\sum_{ 1< m < k} \frac{ \ep_m \ep_k} {12} [ (a_k+a_m)^3-2(a_k^3+a_m^3)] \]
and $Li_3 (e^x)$ is defined by
\[Li_3 (e^x)= \sum_{k=1}^{\infty}\frac { e^{kx}}{ k^3},\]
which has the properities
\[ Li_3^{''}(e^x)= -\ln (1-e^x), \quad Li_3^{'''}(e^x)=coth x= \frac{e^x + e^{-x}}{e^x - e^{-x}}.\]
\indent Using the different row vectors of \eqref{wate}, we can obtain different
 flat coordinate systems  and then get different free energies using the Legendre transformation.

\section{Higher-order Symmetries and conservational Laws}
 In this section, we investigate the symmetries and the conserved densities of waterbag model involving
 quasi-rational function. Quasi-rational means rational with respect to higher derivatives
 . This will generalize the results in \cite{ON} \\
 \indent We start with the recursion operator \eqref{op}. it can be seen that
\begin{equation}
\hat R=J_1 R^{-1} J_1^{-1} =\pa_x (W_x)^{-1}  \label{st}
\end{equation}
is the Sheftel-Teshkov recursion operator \cite{pa2}. And the bi-Hamiltonian structure
\eqref{bi} can also be written as ($n \geq 2$)
\begin{eqnarray}
 \left[\begin{array}{c}
p^1 \\
p^2 \\
\vdots \\
 p^M
\end{array} \right]_{t_n}
&=& \frac{(-1)}{n}J_1
\left[\begin{array}{c}
\frac{\pa h_{n+1} }{\pa  p^1} \\
\frac{\pa h_{n+1} }{\pa  p^2}\\
\vdots \\
\frac{\pa h_{n+1} }{\pa  p^M}
\end{array} \right]    \no \\
  &=&\frac{(-1)}{n(n+1)(n+2)} \hat J_2
\left[\begin{array}{c}
\frac{\pa h_{n+3} }{\pa  p^1} \\
\frac{\pa h_{n+3} }{\pa  p^2}\\
\vdots \\
\frac{\pa h_{n+3} }{\pa  p^M}
\end{array} \right], \label{bi2}
\end{eqnarray}
the Hamiltonian operator $ \hat J_2=\hat R^2  J_1$ being third order. Also,
\begin{equation}
 \hat R  \left[\begin{array}{c}
\frac{\pa h_{n} }{\pa  p^1} \\
\frac{\pa h_{n} }{\pa  p^2}\\
\vdots \\
\frac{\pa h_{n} }{\pa  p^M}
\end{array} \right]=(n-1)
\left[\begin{array}{c}
\frac{\pa h_{n-1} }{\pa  p^1} \\
\frac{\pa h_{n-1} }{\pa  p^2}\\
\vdots \\
\frac{\pa h_{n-1} }{\pa  p^M}
\end{array} \right]. \label{rec4}
\end{equation}
Next, we can express  \eqref{pr} as
\begin{equation}
\left[\begin{array}{c}
p^1 \\
p^2 \\
\vdots \\
 p^M
\end{array} \right]_{y}= \mathbb{H}\left[\begin{array}{c}
p^1 \\
p^2 \\
\vdots \\
 p^M
\end{array} \right]_{x}, \label{pr2}
\end{equation}
where
\[ \mathbb{H}= \left[\begin{array}{ccccc}
p^1+ \ep_1 & \ep_2 & \ep_3 & \ldots  & \ep_M \\
\ep_1 & p^2+\ep_2 & \ep_3 & \ldots & \ep_M \\
\ep_1 & \ep_2 & p^3+\ep_3 & \ldots & \ep_M \\
\vdots & \vdots & \vdots & \vdots & \ep_M  \\
\ep_1 & \ep_2 & \ep_3 & \ldots & p^M+\ep_M
\end{array} \right].     \]
We notice that $(\pa_y -\pa_x \mathbb{H})$ is the Frechet's derivative operator
for the system \eqref{pr2}. It is not difficult to see that if
\[\pa_{\tau} \vec{p}= \vec{\mathbb{Q}}(\vec{p},\vec{p}_x, \vec{p}_{xx}, \cdots)\]
is commuting flow with \eqref{pr2}, then we
have
\begin{equation}
\pa_y \vec{\mathbb{Q}}=\pa_x( \mathbb{H} \vec{\mathbb{Q}}). \label{com}
\end{equation}
\begin{lemma}: Let $W$ be defined in \eqref{wate}. Then \\
(1) $\pa_y W=\mathbb{H} \pa_x W$ ;\\
(2) $\mathbb{H}\pa_x W= (\pa_x W) \mathbb{H}$.
\end{lemma}
\begin{proof}
Direct computations.
\end{proof}
\begin{theorem}
The recursion operator $\hat R$ satisfies the Lax representation
\begin{equation}
\frac{ \pa \hat R}{\pa y} =[\pa_x \mathbb{H}, \hat R ]. \label{lax}
\end{equation}
\end{theorem}
\begin{proof}
Using the lemma, we get
\begin{eqnarray*}
\frac{\pa \hat R}{\pa y} &=& \pa_x \frac{\pa (W_x)^{-1}}{\pa y}=- \pa_x [(W_x)^{-1}W_{xy} (W_x)^{-1}] \\
&=& -\pa_x (W_x)^{-1}[ \mathbb{H}_x W_x +\mathbb{H}W_{xx}] (W_x)^{-1} \\
&=& -\pa_x (W_x)^{-1} \mathbb{H}_x -\pa_x (W_x)^{-1}\mathbb{H}W_{xx} (W_x)^{-1} \\
&=& -\pa_x (W_x)^{-1} \mathbb{H}_x -\pa_x \mathbb{H}(W_x)^{-1}W_{xx} (W_x)^{-1} \\
&=&-\pa_x (W_x)^{-1} \mathbb{H}_x -\pa_x \mathbb{H}[-\pa_x (W_x)^{-1}+(W_x)^{-1}\pa_x]\\
&=&-\pa_x (W_x)^{-1} \mathbb{H}_x +\pa_x \mathbb{H}\pa_x (W_x)^{-1}
-\pa_x \mathbb{H}(W_x)^{-1}\pa_x \\
&=&-\pa_x (W_x)^{-1} \mathbb{H}_x +\pa_x \mathbb{H}\pa_x (W_x)^{-1}
-\pa_x (W_x)^{-1}\mathbb{H}\pa_x \\
&=&\pa_x \mathbb{H}\pa_x (W_x)^{-1}-\pa_x (W_x)^{-1}\pa_x \mathbb{H} \\
&=& [\pa_x \mathbb{H},\pa_x (W_x)^{-1}]=[\pa_x \mathbb{H},\hat R].
\end{eqnarray*}
\end{proof}
Hence from the theorem one knows that the Frechet's derivative operator and recursion
 operator commute,i.e.,
\[(\pa_y-\pa_x \mathbb{H} ) \hat R=\hat R (\pa_y-\pa_x \mathbb{H} ).\]
Morever, if we let ($\mathbb{H}_2=\mathbb{H}$)
\[\left[\begin{array}{c}
p^1 \\
p^2 \\
\vdots \\
 p^M
\end{array} \right]_{t_n}=\frac{(-1)}{n}J_1
\left[\begin{array}{c}
\frac{\pa h_{n+1} }{\pa  p^1} \\
\frac{\pa h_{n+1} }{\pa  p^2}\\
\vdots \\
\frac{\pa h_{n+1} }{\pa  p^M}
\end{array} \right]= \mathbb{H}_n \left[\begin{array}{c}
p^1 \\
p^2 \\
\vdots \\
 p^M
\end{array} \right]_{x}, \]
where
\[ \mathbb{H}_n=\frac{1}{n} \left[\begin{array}{ccccc}
\frac{1}{\ep_1}\frac{\pa^2 h_{n+1}}{\pa p^1 \pa p^1} & \frac{1}{\ep_1}\frac{\pa^2 h_{n+1}}
{\pa p^1 \pa p^2}  & \frac{1}{\ep_1}\frac{\pa^2 h_{n+1}}{\pa p^1 \pa p^3}
 & \ldots  & \frac{1}{\ep_1}\frac{\pa^2 h_{n+1}}{\pa p^1 \pa p^M} \\
\frac{1}{\ep_2}\frac{\pa^2 h_{n+1}}{\pa p^2 \pa p^1} & \frac{1}{\ep_2}\frac{\pa^2 h_{n+1}}
{\pa p^2 \pa p^2}  & \frac{1}{\ep_2}\frac{\pa^2 h_{n+1}}{\pa p^2 \pa p^3}
 & \ldots  & \frac{1}{\ep_2}\frac{\pa^2 h_{n+1}}{\pa p^2  \pa p^M} \\
\frac{1}{\ep_3}\frac{\pa^2 h_{n+1}}{\pa p^3 \pa p^1} & \frac{1}{\ep_3}\frac{\pa^2 h_{n+1}}
{\pa p^3  \pa p^2}  & \frac{1}{\ep_3}\frac{\pa^2 h_{n+1}}{\pa p^3 \pa p^3}
 & \ldots  & \frac{1}{\ep_3}\frac{\pa^2 h_{n+1}}{\pa p^3  \pa p^M} \\
\vdots & \vdots & \vdots & \vdots & \vdots \\
\frac{1}{\ep_M}\frac{\pa^2 h_{n+1}}{\pa p^M \pa p^1} & \frac{1}{\ep_M}\frac{\pa^2 h_{n+1}}
{\pa p^M  \pa p^2}  & \frac{1}{\ep_M}\frac{\pa^2 h_{n+1}}{\pa p^M \pa p^3}
 & \ldots  & \frac{1}{\ep_M}\frac{\pa^2 h_{n+1}}{\pa p^M \pa p^M} \\
\end{array} \right], \]
then from \eqref{rec4} we also yield
\begin{equation}
(\pa_{t_n}-\pa_x \mathbb{H}_n ) \hat R=\hat R (\pa_{t_n}-\pa_x \mathbb{H}_n ). \label{gen}
\end{equation}
Therefore we obtain  the following
\begin{theorem}
 Let $h_{n+1}$ be defined by \eqref{co} and $\mathbb{Q}_m$
be the flows defined by
\[ \vec{p}_{\tau_m}= \left[\begin{array}{c}
p^1 \\
p^2 \\
\vdots \\
 p^M
\end{array} \right]_{\tau_m}
=\mathbb{Q}_m=\hat R^m
\left[\begin{array}{c}
xp^1_x \\
xp^2_x \\
\vdots \\
 xp^M_x
\end{array} \right]=\hat R^{m-1}
\left[\begin{array}{c}
1 \\
1 \\
\vdots \\
1
\end{array} \right] \]
Then $ \mathbb{Q}_m$ is a commuting flow with $ \vec{p}_{t_n}$, i.e.,
$\vec{p}_{t_n t_{\tau_m}}=\vec{p}_{ t_{\tau_m} t_n}$  provided $m \geq n$
\end{theorem}
\begin{proof}
Let's denote
\[ x \vec{p}_x=\left[\begin{array}{c}
xp^1_x \\
xp^2_x \\
\vdots \\
 xp^M_x
\end{array} \right].  \]
(I)
Firstly, one  proves
\[ \hat R (x \vec{p}_x)=\left[\begin{array}{c}
1 \\
1 \\
\vdots \\
1
\end{array} \right] \]
or  \[x \vec{p}_x
=\hat R^{-1} \left[\begin{array}{c}
1 \\
1 \\
\vdots \\
1
\end{array} \right] =W_x \left[\begin{array}{c}
x \\
x \\
\vdots \\
x
\end{array} \right],\]
where $W$ is defined by \eqref{wate}. A direct computation
can obtain this. \\
\\
(II)
Secondly, using \eqref{com} and \eqref{gen}, we have
\begin{eqnarray*}
(\pa_{t_n}-\pa_x \mathbb{H}_n )\mathbb{Q}_m &=& (\pa_{t_n}-\pa_x \mathbb{H}_n )\hat R^m
(x \vec{p}_x)   \\
&=& \hat R^m (\pa_{t_n}-\pa_x \mathbb{H}_n )
(x \vec{p}_x)\\
&=& \hat R^m [ x \pa_x (\vec{p}_{t_n }-\mathbb{H}_n \vec{p}_x)- \mathbb{H}_n \vec{p}_x ] \\
&=& -\hat R^m ( \vec{p}_{t_n})= -n(n-1)(n-2) \cdots (n-m+1) \vec{p}_{n-m},
\end{eqnarray*}
which vanishes if $ m \geq n $ by \eqref{rec4}. \\
\end{proof}
According to the bi-Hamiltonian theory \cite{Ma,Pe}, we can also
express $\vec{p}_{\tau_{2m+1}}$ as
\begin{eqnarray}
\left[\begin{array}{c}
p^1 \\
p^2 \\
\vdots \\
 p^M
\end{array} \right]_{\tau_{2m+1}}&=& \hat R^{2m+1} \left[\begin{array}{c}
xp^1_x\\
xp^2_x \\
\vdots \\
 xp^M_x
\end{array} \right] \no \\
&=& (\hat R)^{2k} J_1 \left[\begin{array}{c}
\frac{\pa \hat  h_{m-k} }{\pa  p^1} \\
\frac{\pa \hat h_{m-k} }{\pa  p^2}\\
\vdots \\
\frac{\pa \hat h_{m-k} }{\pa  p^M}
\end{array} \right], \quad -\infty < k \leq m, \label{bi3}
\end{eqnarray}
where $\hat h_{m}'s$ are Hamiltonian densities, $m=1,2 \cdots,$ with $m$ indicating the
order of derivatives on which they depend, and
\[\hat h_0= -x (\sum_{i=1}^M \ep_i p^i)=-x h_1. \]
We notice that the Hamiltonian densities $\hat h_m$ for $m \geq 1$  are rational functions
in derivatives and can be obtained using the method described in p.69 of \cite{do}. But the
computation is involved and we don't go further here. One also remarks that  the $\hat h_0$ is
\textit {not}
conserved density of \eqref{pr}.  \\
\indent From the theorem, one knows that $\vec{p}_{\tau_{2m+1}}$ commutes with
 $\vec{p}_{t_n}$ provided $2m+1 \geq n $.  Inspired by \cite{ON}(see also \cite{Sh,ts}),
  we have the following \\
 \textbf{ Conjecture:} $\hat h_m $ are conserved densities of $\vec{p}_{t_n}$
  provided $2m+1 \geq n $. \\
\indent In particular, when $n=2$, we get that for all $\hat h_m$, $m=1,2, \cdots, $ they are
conserved densities of \eqref{pr}. \\
\textbf{ Remark:} The Riemann invariants of \eqref{pr} are
\[ \lambda_i= \lambda (u_i),\]
where
\[ \frac{d \lambda}{d z}|_{z=u_i}=0 \]
and the associated Lame cofficients $\Xi_i$ are defined by
\[ \Xi_i^2 ( \vec{\lambda}) =Res_{u_i} \frac{ (d z)^2}{d \lambda}, \]
where $ \vec{\lambda}=(\lambda_1, \lambda_2, \cdots, \lambda_M )$. Then it's shown that
\eqref{pr} has the conserved  density \cite{ts}
\[ h( \vec{\lambda}, \vec{\lambda}_x)= \sum_{k=1}^M \frac{ \Xi_i^2}{\lambda_{i,x}}.\]
One can believe the following identity
\[ \hat h_1 = h( \vec{\lambda}, \vec{\lambda}_x),\]
(up to some scaling)but a  proof is still unknown.
\section{Concluding Remarks}
We establish the bi-Hamiltonian structure \eqref{bi2} (or \eqref{bi}) of the waterbag model
\eqref{wat} using the free energy \eqref{fre} ( or Theorem 2.1)  of topological field theory
associated with it. It turns out that the bi-Hamiltonian structures consists of first-order
and third-order Hamiltonian operators when compared with the compatible Dubrovon-Novikov
Hamiltonian operators \cite{Dn,du1}. The reason is that the waterbag model \eqref{wat} has no Euler
vector field or the free energy \eqref{fre} has no quasi-homogeneous property. In
this situation, it's shown that there is compatible Hamiltonian operators
of first order and third order using such free energy \cite{Mo2,Mo3,pa2}. On the other hand, since the
recursion operator \eqref{st} is local, it's natural to think about the rational
symmetries of waterbag, which tries to generalize the results in \cite{ON} to n-component case.
We conjecture that $\hat h_m$ ($ m \geq 1$) is a conserved density of quasi-rational
function for the hierarchy  under the constraint $2m+1 \geq n$.  These conserved densities
are related to the degenerate Lagrangian representations in flat coordinates  for the  hierarchy of waterbag
model (see \cite{MO,NP}).
In Riemann's invariants, these Lagrangian representations are investigate in \cite{pa2}.  \\
\indent One believes that these results can be generalized to waterbag model of dToda \cite{bk}.
But the computation are more involved and should be addressed elsewhere. \\
{\bf Acknowledgments\/} \\
The author is grateful to Professor Leonid V. Bogdanov's suggestions and  Professor
 Maxim V.Pavlov for his useful discussions
on the paper \cite{pa2}. He also thanks the organizers  of the
Workshop " Mathematical Analysis"( July 2006) in National Tai-Tung  University for their
warm hospitality,
where  part of the work is finished.\\
\indent The work is supported by the National Science Council for the Taiwan-Russia
 joint project under grant no. NSC 95-2923-M-606-001-MY3.

\end{document}